\documentclass[twocolumn,noshowpacs,preprintnumbers,amsmath,amssymb,prl,aps,superscriptaddress]{revtex4}
\usepackage{graphicx}
\usepackage{amsmath}
\usepackage{verbatim}

\newcommand{\ket}[1]{\left\lvert #1 \right\rangle}
\newcommand{\bra}[1]{\left\langle #1 \right\rvert}
\newcommand{\FidL}{F_{\mathrm{L}}}
\newcommand{\FidTQ}{F_{\mathrm{3Q}}}

\newcommand{\Xb}{X_{\mathrm{b}}}
\newcommand{\Xt}{X_{\mathrm{t}}}
\newcommand{\Xm}{X_{\mathrm{m}}}
\newcommand{\Zb}{Z_{\mathrm{b}}}
\newcommand{\Zt}{Z_{\mathrm{t}}}
\newcommand{\Zm}{Z_{\mathrm{m}}}
\newcommand{\Dt}{D_\mathrm{t}}
\newcommand{\Dm}{D_\mathrm{m}}
\newcommand{\Db}{D_\mathrm{b}}
\newcommand{\At}{A_\mathrm{t}}
\newcommand{\Ab}{A_\mathrm{b}}
\newcommand{\Bt}{B_\mathrm{t}}
\newcommand{\Bb}{B_\mathrm{b}}
\newcommand{\MHz}{\mathrm{MHz}}
\newcommand{\GHz}{\mathrm{GHz}}
\newcommand{\Pdp}{P_\mathrm{t}P_\mathrm{b}}

\newcommand{\Pt}{P_\mathrm{t}}
\newcommand{\Pb}{P_\mathrm{b}}

\newcommand{\Dmaj}{\hat{D}}
\newcommand{\abs}[1]{\left| #1 \right|}
\newcommand{\perr}{p_{\mathrm{err}}}
\newcommand{\tr}{\text{Tr}}
\newcommand{\mTorr}{\mathrm{mTorr}}
\newcommand{\nm}{\mathrm{nm}}
\newcommand{\K}{\mathrm{K}}
\newcommand{\cm}{\mathrm{cm}}
\newcommand{\um}{\mu \mathrm{m}}
\newcommand{\degC}{^{\circ}\mathrm{C}}
\newcommand{\mins}{\mathrm{min}}
\newcommand{\mm}{\mathrm{mm}}
\newcommand{\mK}{\mathrm{mK}}
\newcommand{\ns}{\mathrm{ns}}
\newcommand{\Vtop}{V_\mathrm{t}}
\newcommand{\Vbottom}{V_\mathrm{b}}
\newcommand{\Hethree}{^3\mathrm{He}}
\newcommand{\Hefour}{^4\mathrm{He}}

\newcommand{\ketsub}[2]{\ket{#1_{\mathrm{#2}}}}
\newcommand{\brasub}[2]{\bra{#1_{\mathrm{#2}}}}
\newcommand{\kettmb}[1]{\ket{#1_{\mathrm{t}}#1_{\mathrm{m}}#1_{\mathrm{b}}}}

\begin{document}
\title{Detecting bit-flip errors in a logical qubit using stabilizer measurements}

\author{D.~Rist\`e\footnote{equal contribution.}}
\author{S.~Poletto$^\ast$}
\affiliation{QuTech and Kavli Institute of Nanoscience, Delft University of Technology, P.O. Box 5046, 2600 GA Delft, The Netherlands}
\author{M.-Z.~Huang$^\ast$}
\affiliation{QuTech and Kavli Institute of Nanoscience, Delft University of Technology, P.O. Box 5046, 2600 GA Delft, The Netherlands}
\affiliation{Huygens-Kamerlingh Onnes Laboratory, Leiden Institute of Physics, Leiden University, P.O. Box 9504, 2300 RA Leiden, The Netherlands}
\author{A.~Bruno}
\affiliation{QuTech and Kavli Institute of Nanoscience, Delft University of Technology, P.O. Box 5046, 2600 GA Delft, The Netherlands}
\author{V.~Vesterinen}\altaffiliation[Present address: ]{VTT Technical Research Centre of Finland, P.O. Box 1000, 02044 VTT, Finland.}
\author{O.-P.~Saira}\altaffiliation[Present address: ]{Low Temperature Laboratory (OVLL), Aalto University, P.O. Box 15100, FI-00076 Aalto, Finland.}
\author{L.~DiCarlo}
\affiliation{QuTech and Kavli Institute of Nanoscience, Delft University of Technology, P.O. Box 5046, 2600 GA Delft, The Netherlands}

\date{\today}

\maketitle

\textbf{Quantum data is susceptible to decoherence induced by the environment and to errors in the hardware processing it. A future fault-tolerant quantum computer will use quantum error correction (QEC) to actively protect against both.
In the smallest QEC codes~\cite{Shor95,Calderbank96,Bennett96,Laflamme96,Steane96}, the information in one logical qubit is encoded in a two-dimensional subspace of a larger Hilbert space of multiple physical qubits.
For each code, a set of non-demolition multi-qubit measurements, termed stabilizers, can discretize and signal physical qubit errors
without collapsing the encoded information.
Experimental demonstrations of QEC to date, using nuclear magnetic resonance~\cite{Cory98}, trapped ions~\cite{Chiaverini04,Schindler11}, photons~\cite{Pittman05}, superconducting qubits~\cite{Reed12}, and NV centers in diamond~\cite{Waldherr14,Taminiau14}, have circumvented stabilizers at the cost of decoding at the end of a QEC cycle.
This decoding leaves the quantum information vulnerable to physical qubit errors until re-encoding, violating a basic requirement for fault tolerance. Using a five-qubit superconducting processor, we realize the two parity measurements comprising the stabilizers of the three-qubit repetition code~\cite{Nielsen00} protecting one logical qubit from physical bit-flip errors. We construct these stabilizers as parallelized indirect measurements using ancillary qubits, and evidence their non-demolition character by generating three-qubit entanglement from superposition states. We demonstrate stabilizer-based quantum error detection (QED) by subjecting a logical qubit to coherent and incoherent bit-flip errors on its constituent physical qubits.  While increased physical qubit coherence times and shorter QED blocks are required to actively safeguard quantum information, this demonstration is a critical step toward larger codes based on multiple parity measurements.
}

\begin{figure}[t]
\includegraphics[width=\columnwidth]{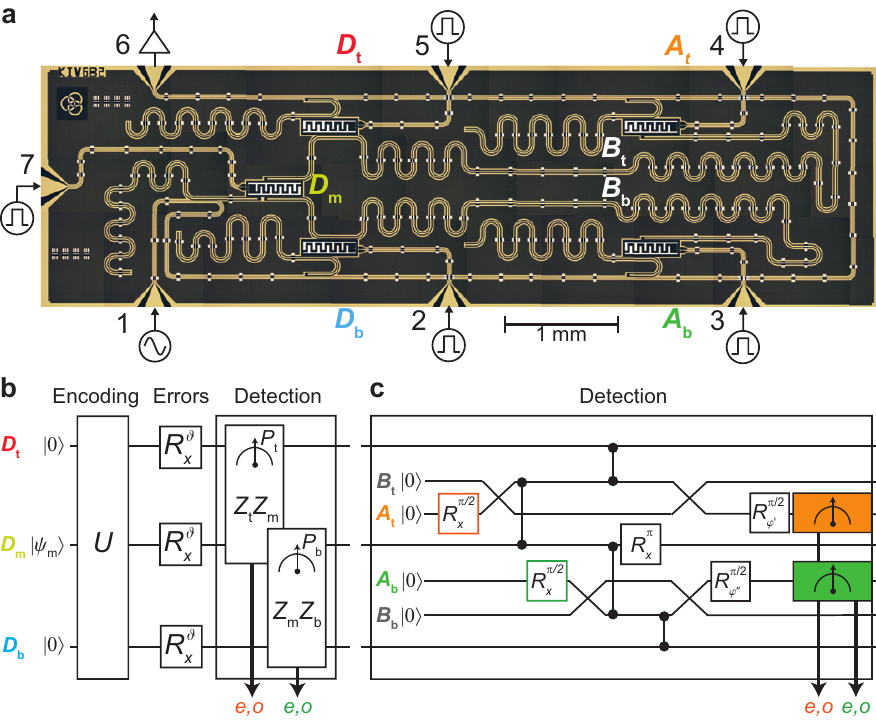}
\caption{\textbf{Quantum processor and gate sequence for implementing and characterizing bit-flip QED by stabilizer measurements.}
	\textbf{a,} Photograph of the processor showing the position and interconnections of data qubits ($\Dt$, $\Dm$, $\Db$), ancilla qubits ($\At$, $\Ab$), buses ($\Bt$, $\Bb$), and
	dedicated readout resonators. These resonators couple to one common feedline to which all readout and microwave control pulses are applied~\cite{Jerger12}.
	Flux-bias lines (ports 2-5, 7) allow control of the qubit transition frequencies on $\ns$ timescale (Extended Data Fig.~1). Details of the processor, including fabrication, parameters and performance benchmarks, are provided in Methods and Extended Data Table~1.
	\textbf{b,} Block diagram for characterizing bit-flip QED by parallelized parity measurements of pairs ($\Dt$, $\Dm$) and ($\Dm$, $\Db$).
	The $\Dm$ state $\ketsub{\psi}{m}=\alpha\ketsub{0}{m}+\beta\ketsub{1}{m}$ is first encoded into the logical qubit state $\ketsub{\psi}{L}=\alpha\kettmb{1}+\beta\kettmb{0}$.
	Coherent or incoherent bit-flip errors are then introduced on data qubits with independent single-bit-flip probability $\perr$.
	Parallelized $\Zt\Zm$ and $\Zm\Zb$ stabilizer measurements discretize these errors, and the two-bit measurement result $\Pdp$ is interpreted as signalling either no error or error on one qubit.
	\textbf{c}, Gate sequence implementing the stabilizer measurements by parallelized interaction with ancilla qubits and projective ancilla measurements.
	Each ancilla is prepared in a superposition state that is transferred to the respective bus with an iSWAP gate (diagonal lines).
	Consecutive CPHASE gates between each bus and the coupled data qubits (vertical lines) encode the data-qubit parity in the quantum phase of the bus superposition state. The final iSWAP transfers this state to the ancilla, and the latter is then projectively measured in the $\ket{\pm}$ basis.
	Halfway through the interaction step, a refocusing $\pi$ pulse is applied to $\Dm$ to reduce inhomogeneous dephasing.}
\end{figure}

A recent roadmap~\cite{Devoret13} for fault-tolerant quantum computing marks a transition from storing quantum data in physical qubits to QEC-protected logical qubits as the fourth of seven development stages.  Following steady improvements in qubit coherence, coherent control, and measurement over 15 years, superconducting quantum circuits are well poised to face this outstanding challenge common to all quantum computing platforms.
Initial experiments using superconducting processors include one round of either bit-flip or phase-flip QEC with decoding~\cite{Reed12}, and the stabilization of one Bell state using dissipation engineering~\cite{Shankar13}. Independent, parallel work~\cite{Corcoles14} demonstrates the detection of general errors on a single Bell state using stabilizer measurements. We demonstrate stabilizer-based QED on the minimal unit of encoded quantum information, a logical qubit, restricting to bit-flip errors.

By analogy to the classical repetition code that maps bit 0 (1) to 000 (111), the quantum version maps the one-qubit state $\alpha \ket{0}+\beta\ket{1}$ to the Greenberger-Horne-Zeilinger-type (GHZ) state $\alpha \kettmb{0}+\beta\kettmb{1}$ of three data qubits (labelled top, middle, and bottom)~\cite{Nielsen00}.
The stabilizers of this code consist of two-qubit parity measurements described by Hermitian operators $\Zt\Zm$ and $\Zm\Zb$.
While GHZ-type states are  eigenstates of both stabilizers with eigenvalue +1, their corruption by a bit-flip error on one  data qubit produces eigenstates with a unique pattern of -1 eigenvalues.
Measuring stabilizers can thus discretize and signal single bit-flip errors without affecting the encoded information (i.e., the probability amplitudes $\alpha$ and $\beta$). Depending on the error signalled, the logical qubit is transformed to an orthogonal two-dimensional subspace. 

This realization of bit-flip QED with stabilizer measurements employs a superconducting quantum processor with 12 quantum elements (Fig.~1a) exploiting resonant and dispersive regimes of circuit quantum electrodynamics~\cite{Blais04}.
Three data transmon qubits ($\Dt$, $\Dm$ and $\Db$) encode the logical qubit. Two ancillary transmons ($\At$ and $\Ab$), two bus resonators ($\Bt$ and $\Bb$), and two dedicated ancilla readout resonators are used for the  stabilizer measurements.
Dedicated readout resonators on data qubits are used to quantify performance (fidelity measures, entanglement witnessing, and  state tomography).
All readout resonators couple to one feedline used for all qubit control and readout pulses.  The feedline output couples to a single amplification chain allowing readout of all qubits by frequency-division multiplexing~\cite{Jerger12}. Ancilla readout fidelity is boosted by a Josephson parametric amplifier (JPA)~\cite{Castellanos-Beltran08} with bandwidth covering both ancilla readout frequencies ($9~\MHz$  apart).

\begin{figure}[t]
\includegraphics[width=\columnwidth]{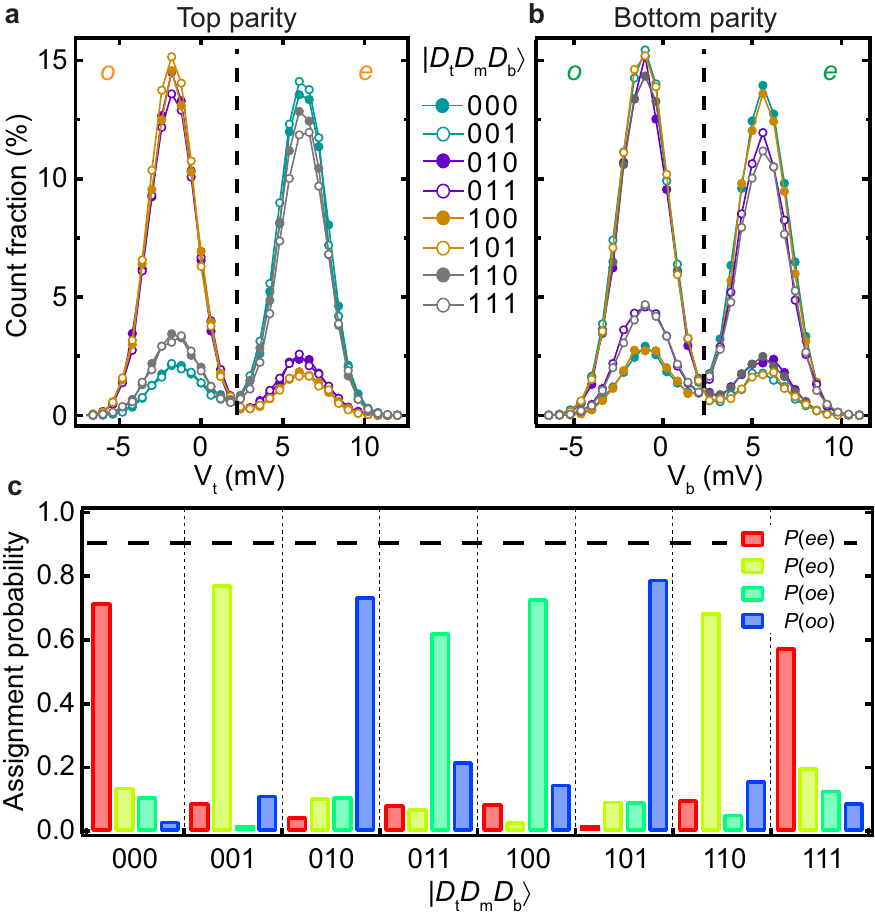}
\caption{\textbf{Characterization of stabilizer measurements.}
	Single-shot histograms for top ($\textbf{a}$) and bottom ($\textbf{b}$) ancilla readout signals $\Vtop$ and $\Vbottom$ at the end of the sequence in Fig.~1c, with data-qubit computational states as input.
	The chosen thresholds for discretization of $\Vtop$ and $\Vbottom$ (dashed vertical lines) maximize the parity assignment fidelities.
	\textbf{c}, Double-parity assignment probabilities for each computational state input. The dashed horizontal line at 0.91 marks the loss of average assignment fidelity exclusively from ancilla readout errors.}
\end{figure}

Building on recent developments~\cite{Saira14,Chow14}, we construct quantum non-demolition  stabilizer measurements in a two-step process combining entanglement with ancilla qubits and their projective measurement. Measuring the stabilizer $\Zt\Zm$ involves an iSWAP gate between $\At$  and $\Bt$, two CPHASE gates between  $\Bt$ and each of $\Dt$ and $\Dm$, and a final iSWAP transferring the $\Bt$ state onto $\At$. These interactions correlate joint states of $\Dt$ and $\Dm$ with even/odd ($e/o$) number of excitations with orthogonal states of $\At$. 
Subsequently, $\At$ is measured by interrogating its dispersively coupled resonator. Conveniently, the interaction and measurement steps needed for both stabilizers can be partially parallelized (Fig.~1c).
(Note that a refocusing $\pi$ pulse is applied to $\Dm$ after its interactions to minimize its inhomogeneous dephasing.)

We begin characterizing these stabilizer measurements by testing their ability to detect the parities of the computational states $\ket{i_{\mathrm{t}} j_{\mathrm{m}} k_{\mathrm{b}}}$, $i,j,k\in\{0,1\}$.
Because all of these states are eigenstates of $\Zt\Zm$ and $\Zm\Zb$, a fixed two-bit measurement outcome $\Pdp\in\{ee,eo,oe,oo\}$ is expected for each one.
Histograms of declared double parities clearly reveal the correlation (Fig.~2).
The average assignment fidelity of $71\%$, defined as the probability of correct double-parity assignment averaged over the eight states, is limited by errors in the interaction step (separate calibrations of ancilla readout errors set a $91\%$ upper bound).

\begin{figure}[t]
\includegraphics[width=\columnwidth]{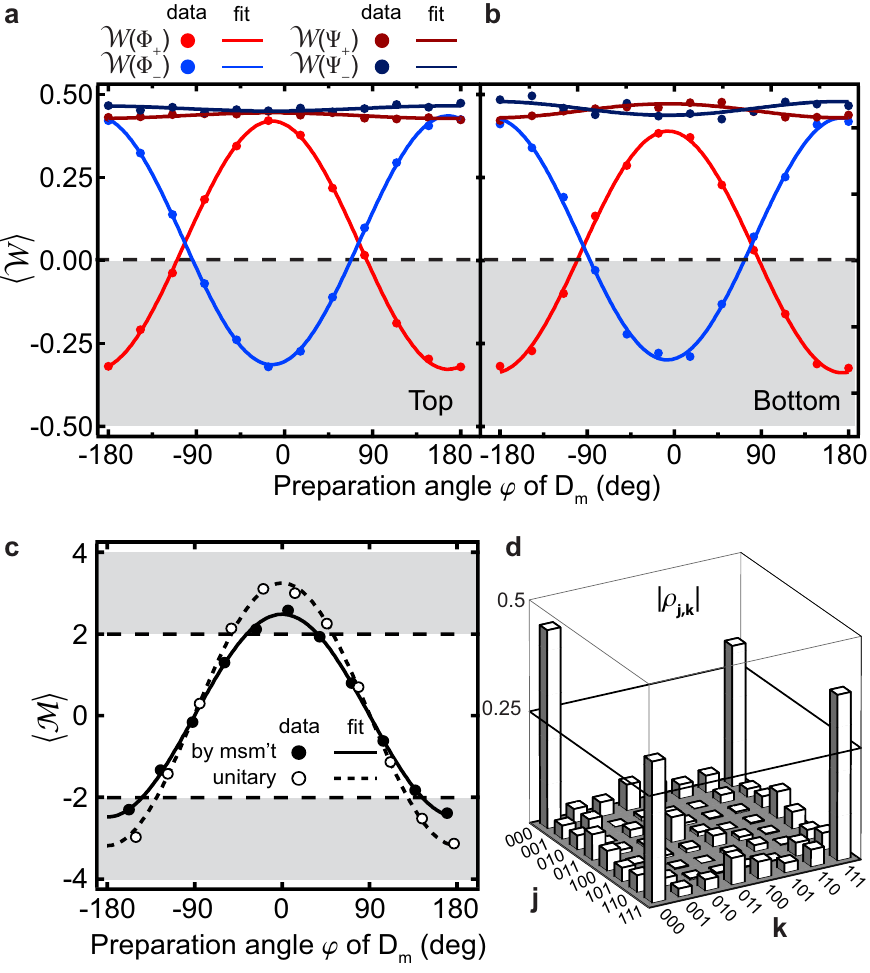}
\caption{\textbf{Generation of two- and three-qubit entanglement by stabilizer measurements.}
	Starting with the data qubits in the state $\ketsub{+}{t}\left(\ketsub{0}{m}+e^{i\varphi}\ketsub{1}{m}\right)\ketsub{+}{b}/\sqrt{2}$, we selectively perform stabilizer measurements by activating the corresponding ancilla (applying initial $\pi/2$ rotation in Fig.~1c).
	\textbf{a, b,} Performing one parity measurement generates entanglement between the paired data qubits.
	Measured average $\langle\mathcal{W}\rangle$ of the four witnesses operators $\mathcal{W}({\Phi_\pm})$ and $\mathcal{W}({\Psi_\pm})$ involving the data qubits paired by activating the top ($\textbf{a}$) or bottom ($\textbf{b}$) ancilla only and postselection on $\Pt =o$ and $\Pb =o$, respectively.
	Entanglement is witnessed whenever $\langle\mathcal{W}\rangle <0$.
	The weak oscillations in $\langle \mathcal{W}({\Psi_\pm})\rangle$ result from false positives, which we have partially reduced here by postselecting more strongly than the threshold maximizing the average parity assignment fidelity.
	A dual witnessing by $\langle\mathcal{W}({\Psi_\pm})\rangle$ is observed by postselection on $e$.
	\textbf{c}, Measured average of the Mermin operator $\mathcal{M}$ with both ancillas activated and data strongly postselected on $\Pdp=oo$ (black circles).
	Three-qubit entanglement is witnessed whenever $\abs{\mathcal{M}}>2$.
	A stronger violation of the Mermin inequality is observed when targeting the GHZ state $\ket{\mathrm{GHZ}}=\left(\kettmb{0}+\kettmb{1}\right)/\sqrt{2}$ using unitary gates only (white circles).
	\textbf{d}, Tomography (absolute value of the density matrix elements) of the $\abs{\mathcal{M}}$-maximizing state generated by double-parity measurement. The fidelity $F=\bra{\mathrm{GHZ}}\rho\ket{\mathrm{GHZ}}$ is $73\%$.
	For comparison, targeting this state with gates achieves $F=82\%$.}
\end{figure}

The next test probes the ability of each stabilizer to discern two-qubit parity subspaces while preserving coherence within each.
Specifically, we target the generation of two- and three-qubit entanglement (2QE and 3QE) via single and double stabilizer measurements on a maximal superposition state.
The gate sequence in Fig.~1c is executed with $\Dt$ and $\Db$ both prepared in $\ket{+}=(\ket{0}+\ket{1})/\sqrt{2}$ and $\Dm$ in $\left(\ket{0}+e^{i\varphi}\ket{1}\right)/\sqrt{2}$.
First, we activate one stabilizer by performing the initial $\pi/2$ rotation only on the corresponding ancilla, and  measure the data-qubit-pair witness operators $\mathcal{W}({\Phi_\pm})=\left( II\mp XX\pm YY-ZZ\right)/4$, $\mathcal{W}({\Psi_\pm})=\left( II\mp XX\mp YY+ZZ\right)/4$~\cite{Horodecki09} based on fidelity to even- and odd-parity Bell states, respectively.
Each of these operators witnesses 2QE whenever the expectation value $\langle \mathcal{W} \rangle <0$. With postselection on result $o$, $\langle \mathcal{W}({\Phi_+})\rangle$ and $\langle \mathcal{W}({\Phi_-})\rangle$ jointly witness 2QE at almost all values of $\varphi$  (Figs.~3a and 3b). 

We continue building multi-qubit entanglement by activating both parity measurements and postselecting on the two-bit result (Figs.~3c, 3d, and Extended Data Fig.~2).
Ideally, $\Pdp=oo$ collapses the maximal superposition onto the GHZ-type state $\ket{\mathrm{GHZ}(\varphi)}=\left(\kettmb{0}+e^{-i\varphi}\kettmb{1}\right)/\sqrt{2}$.
Genuine 3QE is witnessed whenever $\abs{\langle \mathcal{M} \rangle}>2$, where $\mathcal{M}$ is the Mermin operator
$X_\mathrm{t} X_\mathrm{m} X_\mathrm{b} - Y_\mathrm{t} Y_\mathrm{m} X_\mathrm{b} - Y_\mathrm{t} X_\mathrm{m} Y_\mathrm{b} - X_\mathrm{t} Y_\mathrm{m} Y_\mathrm{b}$~\cite{Mermin90}.
With postselection on $\Pdp=oo$, $\langle \mathcal{M} \rangle$ versus $\varphi$ reaches 2.5 (best fit, Fig.~3c).
Full state tomography at the optimal $\varphi$ reveals a fidelity $\bra{\mathrm{GHZ}(0)}\rho\ket{\mathrm{GHZ}(0)}=73\%$ to the ideal GHZ state (Fig.~3d).

This 3QE-by-measurement protocol can also be used to perform the encoding step of bit-flip QEC.
Ideally, the state $\ketsub{+}{t}\left(\alpha \ketsub{0}{m}+\beta \ketsub{1}{m}\right)\ketsub{+}{b}$ is mapped onto $\alpha\kettmb{1}+\beta\kettmb{0}$ up to the transformation $\Xt\Xb$, $\Xt$, $\Xb$, $I$ signalled by $\Pdp=ee$, $eo$, $oe$, $oo$, respectively.
Postselection on $\Pdp=oo$ (Extended Data Fig.~3) encodes with $73\%$ fidelity, averaged over the six cardinal input states of $\Dm$, $\ketsub{\psi^j}{m}\in\left\lbrace\ket{0}, \ket{1}, \ket{\pm}=\left(\ket{0}\pm\ket{1}\right)/\sqrt{2}, \ket{\pm i}=\left(\ket{0}\pm i\ket{1}\right)/\sqrt{2}\right\rbrace $).
For comparison, implementing the standard unitary encoding~\cite{DiCarlo10,Neeley10,Reed12} using our gate toolbox (Extended Data Fig.~4) achieves $82\%$ average fidelity. 

Finally, we use this encoding by gates to demonstrate bit-flip QED by parallelized stabilizer measurements (Fig.~4a).
Bit-flip errors are coherently added via $X$ rotations by an angle $\theta$, yielding a single-qubit bit-flip probability $\perr=\sin^2\left(\theta /2\right)$ (adding incoherent errors at this stage yields very similar results, see Methods and Extended Data Fig.~5).
We consider two scenarios: errors added on only one data qubit (1), and errors added on all three (3).
We first quantify QED performance using the average fidelity $\FidTQ$ to the ideal three-qubit state accounting for the subspace transformation $\hat{C}_{pq}=\Xm,\Xm\Xb,\Xt\Xm,I$ signalled by $\Pdp=ee,eo,oe,oo$ (in order):
\[
\FidTQ = \frac{1}{6} \sum_j \sum_{pq} p_{pq} \brasub{\psi^j}{L}\hat{C}_{pq}\rho(j,pq)\hat{C}^{\dagger}_{pq}\ketsub{\psi^j}{L} \quad \text{(QED)}.
\]
Here, $\ketsub{\psi^j}{L}$ is the ideal encoded cardinal state, $p_{pq}$ is the measured probability of $\Pdp=pq$, and $\rho(j,pq)$ is the experimental $pq$-conditioned density matrix.
The near constancy of $\FidTQ(\perr)$ in scenario (1) and the second-order dependence in (3) (Fig.~4b) reflect the ability of the stabilizers to discretize and signal single-qubit bit-flip errors without decoding. 
 
\begin{figure}
\includegraphics[width=\columnwidth]{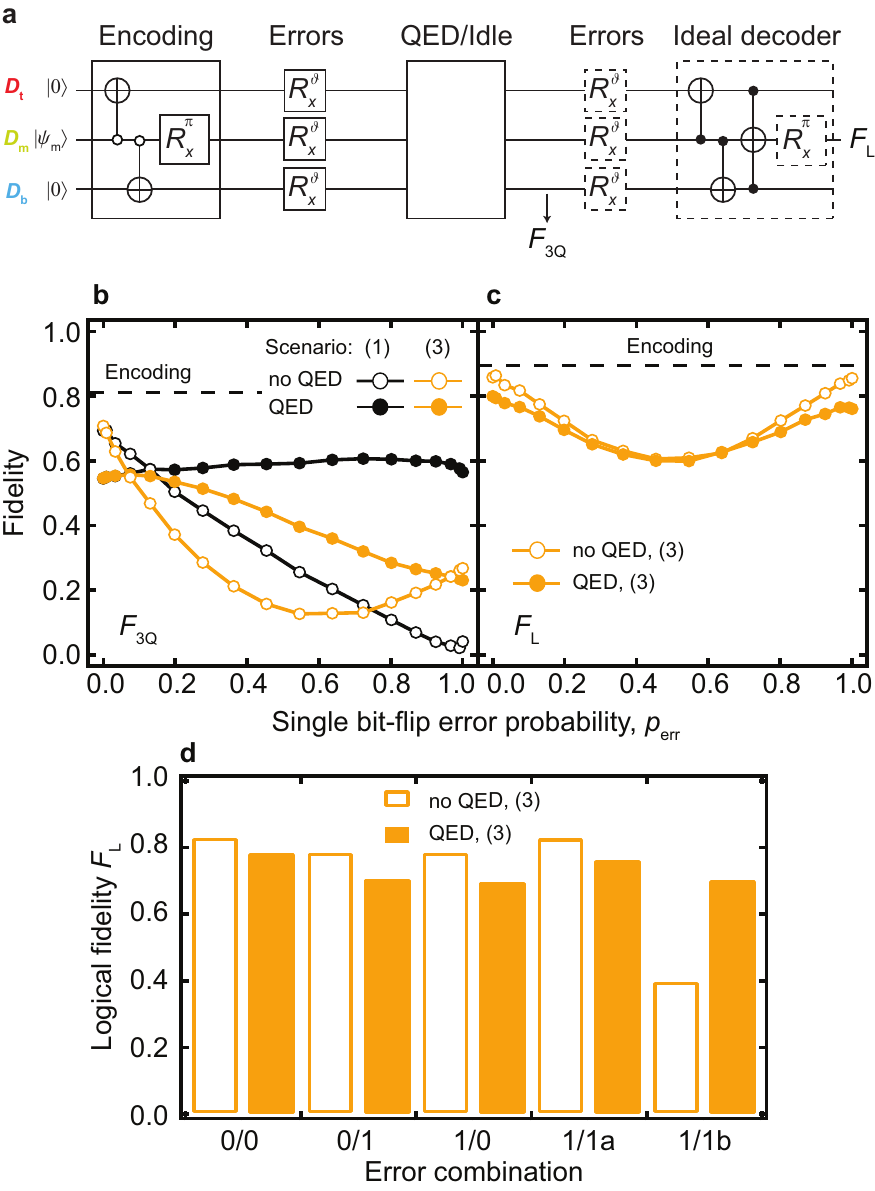}
\caption{\textbf{Detection of bit-flip errors.}
	\textbf{a}, Sequence used to assess performance of bit-flip QED. After encoding by gates, either coherent ($\theta \in [0,\pi]$) or incoherent ($\theta = 0$ or $\pi$) errors are introduced with single-qubit bit-flip probability $\perr$. Next, parallelized stabilizer measurements are either performed or replaced by an equivalent idling period. Partial tomography at this point is used to obtain the three-qubit fidelity $\FidTQ$ and the logical fidelity $\FidL$. The calculation of $\FidL$ assumes incoherent second-round errors with the same $\perr$ and a perfect decoding (dashed boxes).
	\textbf{b}, Three-qubit fidelity $\FidTQ$ as a function of $\perr$ with and without QED under two scenarios: coherent errors applied on $\Dm$ (1) and on all data qubits (3).
	The dashed line indicates the fidelity ceiling imposed by encoding errors.
	\textbf{c},  $\FidL$ as function of $\perr$, obtained from the same data as in b. The individual contributions of the six cardinal states $\ketsub{\psi^j}{m}$ to $\FidTQ$ and $\FidL$ are shown in Extended Data Fig.~7. 
	\textbf{d},  $\FidL$ for all combinations of one and zero incoherent errors before and after QED or idling.
	Error combinations are labelled $m/n$, with $m$ ($n$) the number of errors before (after) QED or idling. 
	The case 1/1 is divided in two: errors on the same data qubit (1/1a) or on different qubits (1/1b).}
\end{figure}

To assess the ability of QED to detect added errors without penalizing for intrinsic decoherence and encoding errors, we compare to $\FidTQ$ with the stabilizer interactions replaced by idling for equal duration (with a refocusing $\Dm$ pulse):
\[
\FidTQ=\frac{1}{6} \sum_j \brasub{\psi^j}{L}\Xm \rho(j) \Xm \ketsub{\psi^j}{L} \quad \text{(no QED)}.
\]
Without QED, one expects a linear decrease in $\FidTQ$ in (1) as one bit flip orthogonally transforms the encoded state. The slight curvature observed reflects residual coherent errors in encoding. The non-monotonicity of $\FidTQ$ in (3) reflects that triple errors perform a logical bit flip, which  leaves $\ket{+_\mathrm{L}}$ and  $\ket{-_\mathrm{L}}$ unchanged. Comparing curves suggests that QED provides net gains for $\perr\gtrsim 15\%$ in (1) and for $\perr\gtrsim 10\%$ in (3) (Fig.~4b).

However, the true merit of QED hinges on the ability to suppress the accumulation of errors. We believe that a  better comparison is the logical state fidelity $\FidL$ following two rounds of errors with QED or idling in between. $\FidL$ is defined as the fidelity to the initial unencoded $\Dm$ state following an ideal decoder $\Dmaj$ (Fig.~4a) that is resilient to a bit-flip error remaining in any one qubit. For example, with QED and a second-round error $\hat{E}$,
\[
\FidL = \frac{1}{6}\sum_{jpq} p_{pq} \brasub{\psi^j}{m}\tr_{\mathrm{t,b}}\!\left[ \Dmaj \hat{E}\hat{C}_{pq}\rho(j,pq)\hat{C}^{\dagger}_{pq}\hat{E}^{\dagger}\Dmaj^{\dagger} \right]\!\ketsub{\psi^j}{m}.
\]
Here we consider scenario (3) and only incoherent second-round errors. We expect QED to win over idling in  select cases, such as single errors on both rounds but on different qubits, all of which we observe (Fig.~4d and also Extended Data Fig.~6). 
Weighing in all possible cases (from 0 to 3 errors in each round) according to their probability, we find that the current fidelity of the stabilizer measurements precludes boosting $\FidL$ by QED at any $\perr$ (Fig.~4c). This stricter comparison sets the benchmark for gauging future improvements in QED.

In summary, we have realized parallel stabilizer measurements with ancillary qubits and used them to perform bit-flip QED in a superconducting circuit. Stabilizer-based QED can detect bit-flip errors on data qubits while maintaining the encoding at the logical level, thus meeting a necessary condition for fault-tolerant quantum computing. Evidently, it remains a priority to extend qubit coherence times and shorten the QED step in order to boost logical fidelity by QED. Future work will also target the completion of several QEC cycles, using digital feedback control~\cite{Riste12b} to correct inferred errors or adapting logical operations in accordance to the subspace transformations signalled by the stabilizer measurements. In the longer term, parallelized ancilla-based parity measurements as demonstrated here may be used to protect a logical qubit against general errors with a Steane~\cite{Steane96,Nigg14} or small surface code~\cite{Tomita14}.
\newpage

\section{Methods}

\textbf{Processor fabrication}.
The integrated circuit is fabricated on a c-plane sapphire substrate.
A NbTiN film ($80~\nm$) is reactively sputtered at $3~\mTorr$ in a 5\% N$_2$ in Ar atmosphere, resulting  in a superconducting critical temperature of $15.5~\K$ and normal-state resistivity of $110~\mu\Omega\cm$.
This film is e-beam patterned using SAL601 resist and etched by SF$_6$/O$_2$ RIE to define all coplanar waveguide structures: feedline, resonators, and flux-bias lines.
The transmon Josephson junctions and shunting interdigitated capacitors are patterned using PMGI/PMMA e-beam lithographed resist and double-angle shadow evaporation of Al with intermediate oxidization.
Air bridges are added to suppress slot-line propagation modes, to connect ground planes, and to allow the crossing of transmission lines (Extended Data Fig.~8).
Bridge fabrication starts with a $6~\um$ thick PMGI layer which is patterned and then reflowed at $220\degC$ for $5~\mins$, producing  a gently arched profile. A second MAA/PMMA resist layer is spun and e-beam patterned to define the bridge geometry. Finally, Ti ($5~\nm$) and Al ($450~\nm$) are e-beam evaporated.
The $2~\mm$ by $7~\mm$ chip is diced and cleaned in $88\degC$ NMP for $30~\mins$.


\textbf{Experimental setup}.
The quantum processor is anchored to the mixing chamber plate of a dilution refrigerator with $15 - 20~\mK$ base temperature.
A detailed schematic of the experimental setup at all temperature stages is shown in Extended Data Fig.~8.
The single coaxial line for readout and microwave control has in-line attenuators and  absorptive low-pass filters providing thermalization, noise reduction, and infrared radiation shielding.
Coaxial lines for flux control are broadband attenuated and bandwidth limited ($1~\GHz$) with reactive and absorptive low-pass filters.


\textbf{Qubit control}.
Most microwave pulses for $X$ and $Y$ qubit rotations have a Gaussian envelope in the main quadrature ($5~\ns$ sigma and $20~\ns$ total duration), and a derivative-of-Gaussian envelope in the other (DRAG pulses~\cite{Motzoi09}).
Wah-Wah pulses~\cite{Vesterinen14} combining DRAG with sideband modulation are used for $\Dt$ and $\Ab$ to avoid leakage in $\Dm$ and $\Db$, respectively.
Taking advantage of the proximity in frequency between $\Dt$ and $\At$, and between $\Dm$ and $\Ab$, we coherently control the five qubits by sideband modulation of three carriers (Extended Data Fig.~8).

Flux pulses for iSWAPs are sudden ($12~\ns$ duration), while those for CPHASEs are mostly fast adiabatic~\cite{Martinis14} ($40~\ns$).
The pulse for CPHASE between $\Dm$ and $\Bt$ is kept sudden ($19~\ns$) to avoid leakage  during the crossing of $\Dm$ through $\Bb$.
Pulse distortion resulting from the flux control bandwidth is minimized by manual optimization of convolution kernels.


\textbf{Qubit readout}.
The five qubits are readout by frequency division multiplexing~\cite{Jerger12}.
The readout pulses for data and ancilla qubits are separately generated by sideband modulation of two carriers.

The amplitude and duration of readout pulses are chosen to maximize assignment fidelity.
$\Dt$, $\Dm$, and $\Db$ readout pulses have $1200$, $1000$, and $700~\ns$ duration, respectively.
The signal-to-noise boost provided by the JPA allows shorter ancilla qubit readouts, $600~\ns$ ($550~\ns$) for $\At$ ($\Ab$).
The amplified feedline output is split and down-converted with two local oscillators.
The two signals are amplified, digitized, demodulated, and integrated to yield one voltage for each qubit measured.
The low crosstalk between the qubit readouts is evidenced by simultaneous measurement immediately following preparation of the 32 combinations of the five qubits in either $\ket{0}$ or $\ket{1}$ (Extended Data Fig.~9).

Using the method of Ref.~\onlinecite{Saira14} based on Hahn echo sequences, we have bound the dephasing of each data qubit induced by the ancilla measurements to less than $1\%$ (data not shown). Since
data-qubit fidelity loss during ancilla measurements is dominated by intrinsic decoherence and our main interest is to quantify the ability of stabilizers to detect the intentionally added errors,
we have opted to advance the data qubit measurements, making them simultaneous to those of ancillas  (Extended Data Fig.~4).


\textbf{Initialization}.
The four qubits $\{\Dt,\Db,\At,\Ab\}$ and two buses $\{\Bt,\Bb\}$ are initialized to their ground state by postselection on six measurements performed before any encoding or manipulation protocol. The buses are initialized by swapping states with their coupled ancilla immediately after initialization of the latter.
$\Dm$ is initialized by swapping its excitation ($\sim 10\%$) with that of $\Bb$ ($\sim 1\%$).
The postselected fraction of runs $(50-60\%)$ have a residual excitation of $1 - 2\%$ in every quantum element.


\textbf{Gate sequence}.
Gates are parallelized as much as possible. We note two important exceptions. Because of frequency crowding and the common feedline, pulses targeting one qubit induce ac Stark shifts on untargeted qubits. We serialize single-qubit control to restrict the effect of these shifts to residual phase rotations on unaddressed qubits. Also, the first iSWAP between $\Bt$ and $\At$ and CPHASE between $\Bt$ and $\Dm$ (Fig.~1c) are applied before populating $\Bb$ to avoid a strong dispersive shift of $\Dm$. All others iSWAPS, CPHASE gates and ancilla measurements are simultaneous.

\textbf{Incoherent errors}.
We have also tested stabilizer-based QED with incoherent first-round errors generated using $\pi$ rotations (Extended Data Fig.~5).
Following encoding of a $\Dm$ cardinal input state $\ketsub{\psi^j}{m}$, we apply the eight combinations of error/no error on the three data qubits. We calculate $\FidTQ$ and $\FidL$ for each combination and weigh by the corresponding probability.

\section{Acknowledgments}
\begin{acknowledgments}
We thank  K.W.~Lehnert for providing the parametric amplifier, D.J.~Thoen and T.M.~Klapwijk for sputtering of NbTiN films, K.M.~Svore,  T.H.~Taminiau,  D.P.~DiVincenzo,  E.~Magesan, and J.M.~Gambetta for fruitful discussions, and N.K.~Langford, G.~de~Lange, L.M.K.~Vandersypen, and R.~Hanson for helpful comments on the manuscript.
We acknowledge funding from the Netherlands Organization for Scientific Research (NWO), the Dutch Organization for Fundamental Research on Matter (FOM), and the EU FP7 project ScaleQIT.
\end{acknowledgments}

\section{Author Contributions}
A.B. fabricated the processor, with design input from O.-P.S. and L.D.C.
O.-P.S. and V.V. performed the initial tune-up.
D.R., M.-Z.H., and S.P. performed measurements and data analysis.
S.P., D.R. and L.D.C. prepared the manuscript with feedback from all other authors.
L.D.C. supervised the project. \\
Correspondence and requests for materials should be addressed to L.D.C. (l.dicarlo@tudelft.nl).

\makeatletter
\renewcommand{\fnum@figure}{Extended Data Table~\thefigure}
\makeatother
\setcounter{figure}{0}

\begin{figure*}[t]
\includegraphics[width=\columnwidth]{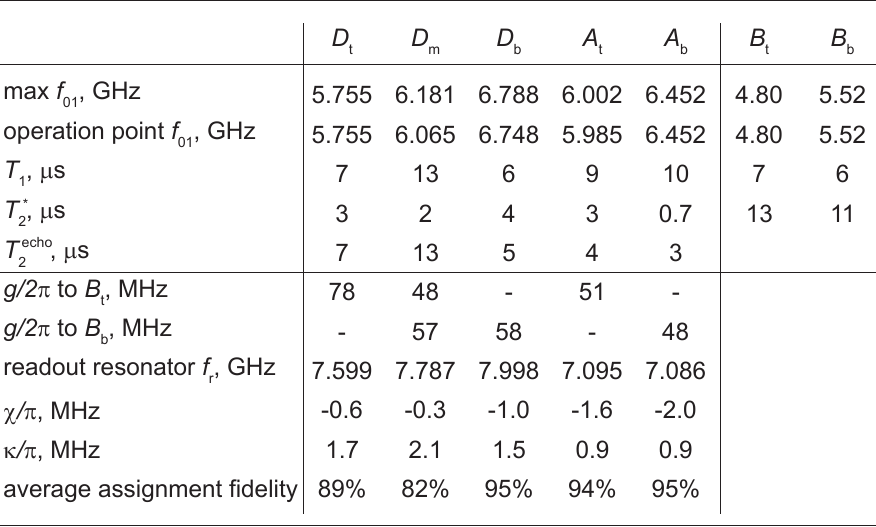}
\caption{\textbf{Summary of the main device parameters.}}
\end{figure*}

\makeatletter
\renewcommand{\fnum@figure}{Extended Data Figure~\thefigure}
\makeatother
\setcounter{figure}{0}

\begin{figure*}[t]
\includegraphics[width=\columnwidth]{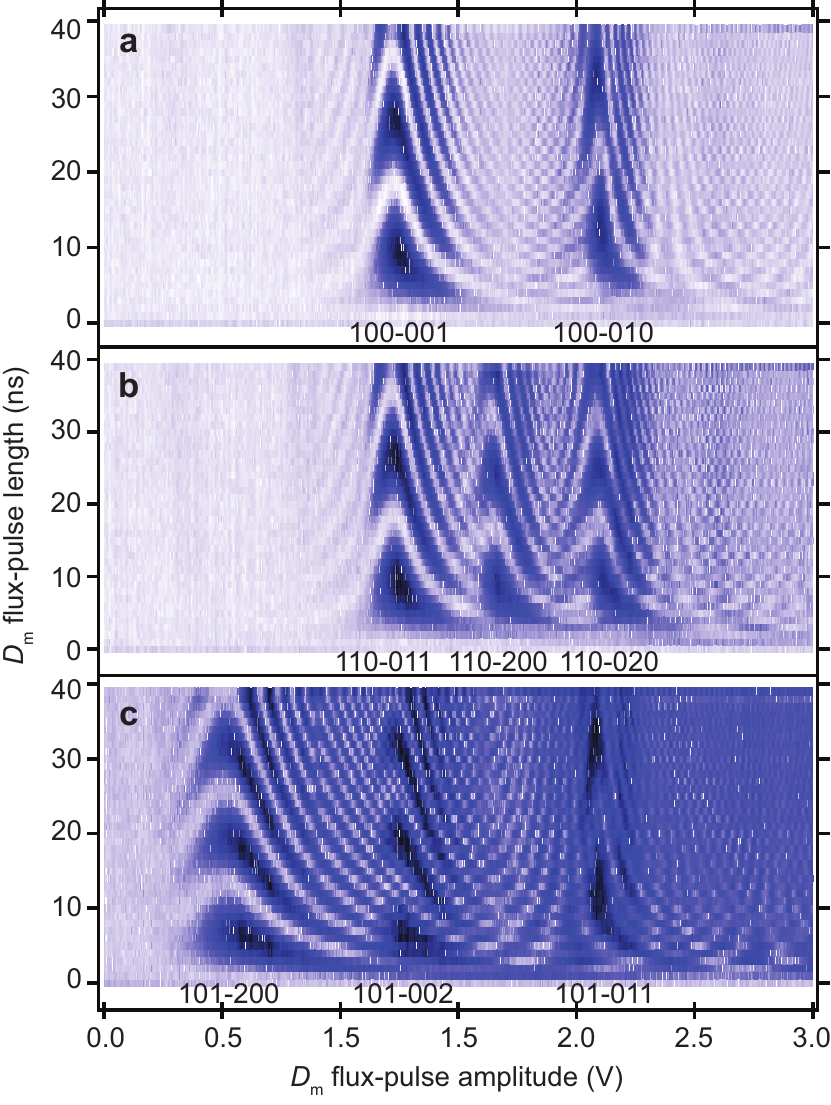}
\caption{\textbf{Vacuum Rabi oscillations between $\Dm$ and the buses.}
	Coherent oscillations between $\Dm$ (initially in $\ket{1}$) and both buses, as a function of flux pulse amplitude and duration. Buses are prepared in $\ket{\Bt \Bb} = \ket{00}$ (\textbf{a}), $\ket{\Bt \Bb} = \ket{10}$ (\textbf{b}), and $\ket{\Bt \Bb} = \ket{01}$ (\textbf{c}). Labels indicate the corresponding transition with notation $\ket{\Dm \Bt \Bb}$.}
\end{figure*}

\begin{figure*}[t]
\includegraphics[width=2\columnwidth]{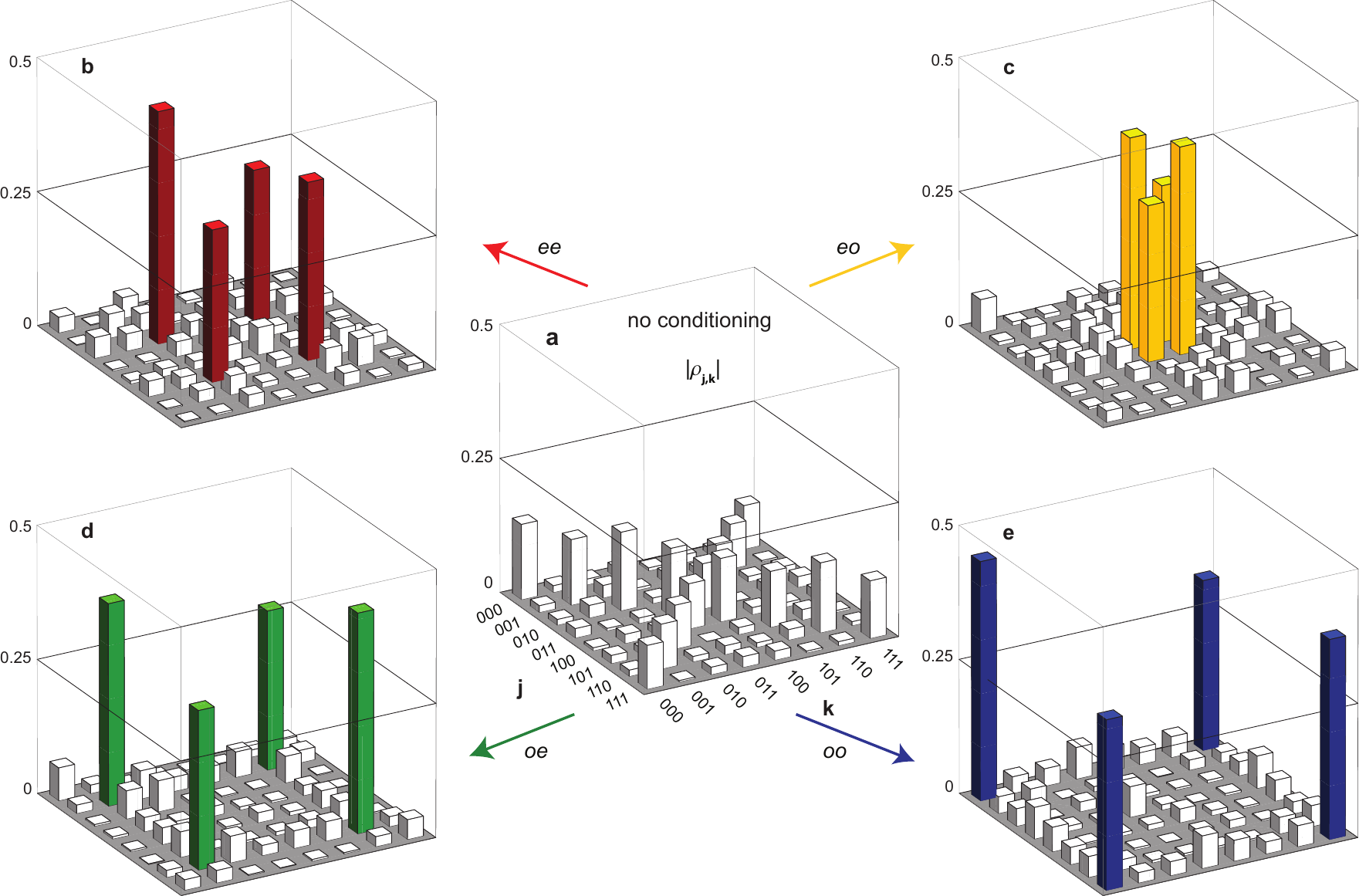}
\caption{\textbf{Three-qubit entanglement by parallelized stabilizer measurements on a maximal superposition state.}
	Density-matrix elements (absolute values) of the states obtained by postselection on different stabilizer measurement results:
	(\textbf{a}) No postselection;
	(\textbf{b}) $\Pdp=ee$, fidelity $\bra{\mathrm{GHZ}}\Xb\Xt\rho\Xb\Xt\ket{\mathrm{GHZ}}=67\%$;
	(\textbf{c}) $\Pdp=eo$, $\bra{\mathrm{GHZ}}\Xt\rho \Xt\ket{\mathrm{GHZ}}=67\%$;
	(\textbf{d}) $\Pdp=oe$, $\bra{\mathrm{GHZ}}\Xb\rho \Xb\ket{\mathrm{GHZ}}=65\%$;
	(\textbf{e}) $\Pdp=oo$, $\bra{\mathrm{GHZ}}\rho\ket{\mathrm{GHZ}}= 68\%$.
	Note that the parities of the final state differ from the detected ones due to the refocusing $\pi$ pulse on $\Dm$.
	In contrast to Fig.~3, conditioning here is performed using the $\Vtop$ ($\Vbottom$) threshold maximizing the top (bottom) parity assignment fidelity.}
\end{figure*}

\begin{figure*}[t]
\includegraphics[width=\columnwidth]{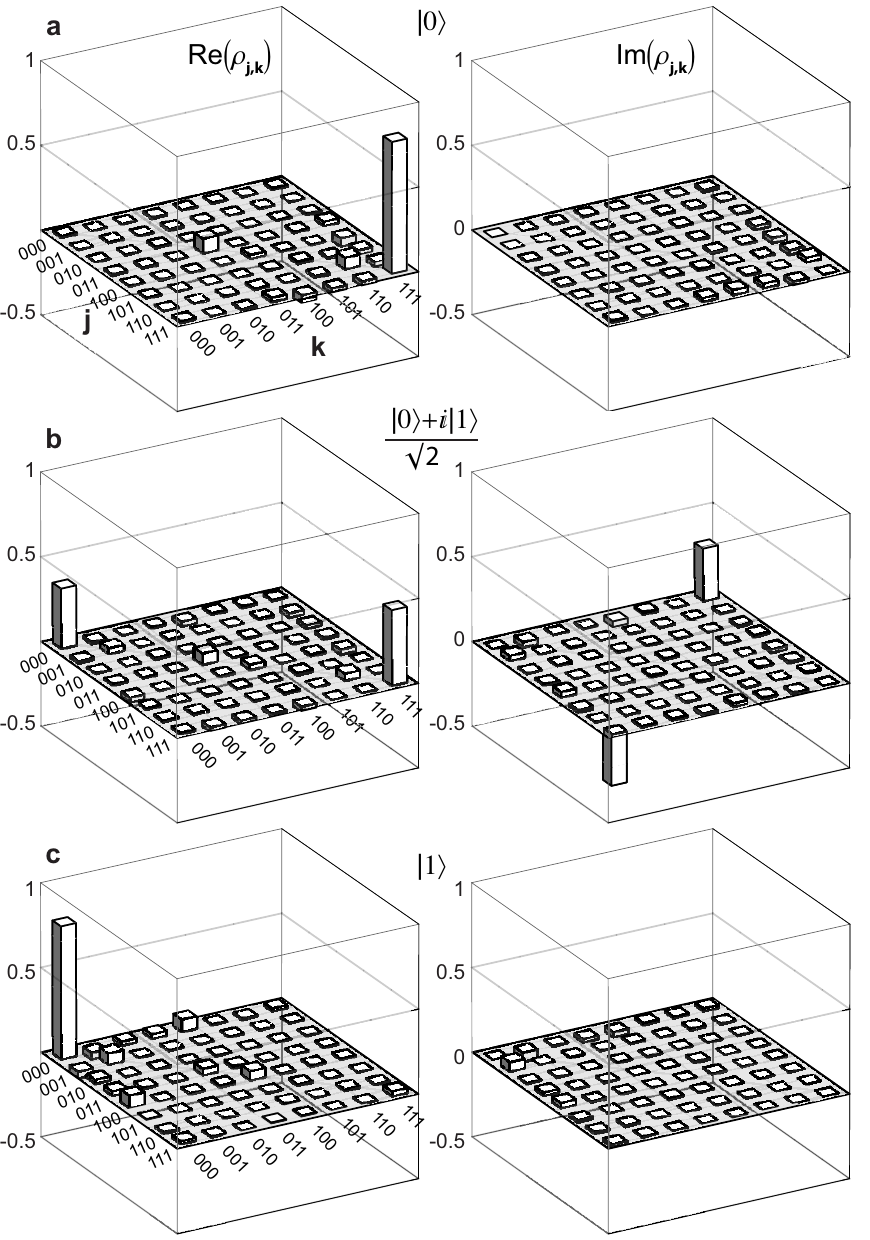}
\caption{\textbf{Encoding by measurement.}
	Density-matrix elements (real and imaginary parts) of the state obtained by stabilizer measurements on the state $\ketsub{+}{t}\ketsub{\psi}{m}\ketsub{+}{b}$ and with strong postselection  on $\Pdp=oo$ (as in Fig.~3), with
	$\ketsub{\psi}{m}=\ket{0}$ (\textbf{a}),
	$\ketsub{\psi}{m}=(\ket{0}+i\ket{1})/\sqrt{2}$ (\textbf{b});
	$\ketsub{\psi}{m}=\ket{1}$ (\textbf{c}).
	Due to the refocusing pulse on $\Dm$, the state $\ket{0}$ ($\ket{1}$) is encoded in $\kettmb{1}$ ($\kettmb{0}$).}
\end{figure*}

\begin{figure*}[t]
\includegraphics[width=2\columnwidth]{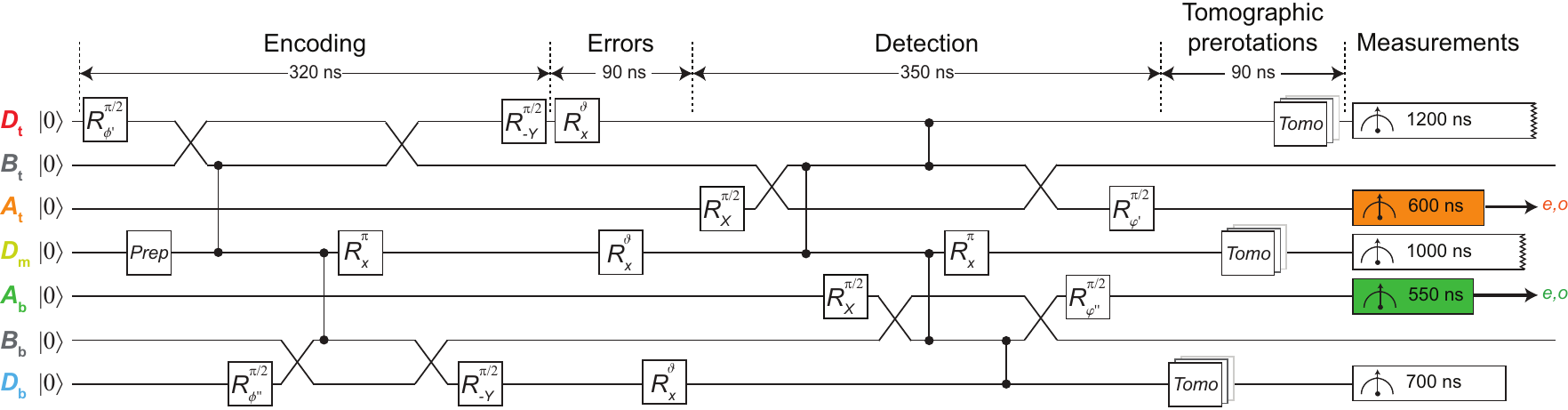}
\caption{\textbf{Quantum circuit for QED characterization.}
	The quantum circuit for QED characterization has six steps: initialization (not shown), encoding, addition of bit-flip errors, detection, tomographic pre-rotation pulses, and measurements.}
\end{figure*}

\begin{figure*}[t]
\includegraphics[width=110mm]{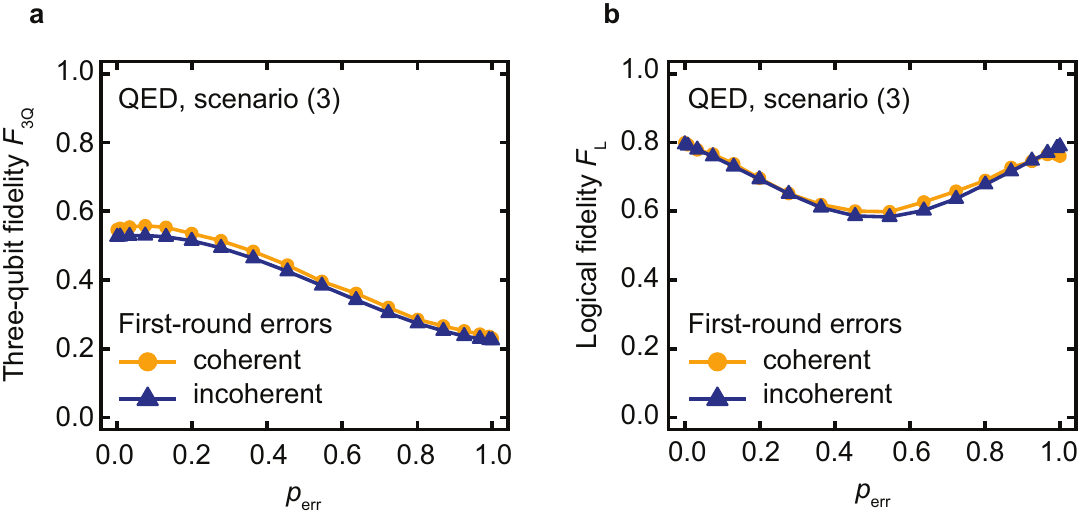}
\caption{\textbf{Comparison between coherent and incoherent added errors.}
	Comparison of fidelities $\FidTQ$ (a) and $\FidL$ (b) for coherent (circles, same data as in Fig.~4b) and incoherent (triangles) errors applied on the first round and for scenario (3) with QED. As expected, the curves closely overlap.}
\end{figure*}

\begin{figure*}[t]
\includegraphics[width=2\columnwidth]{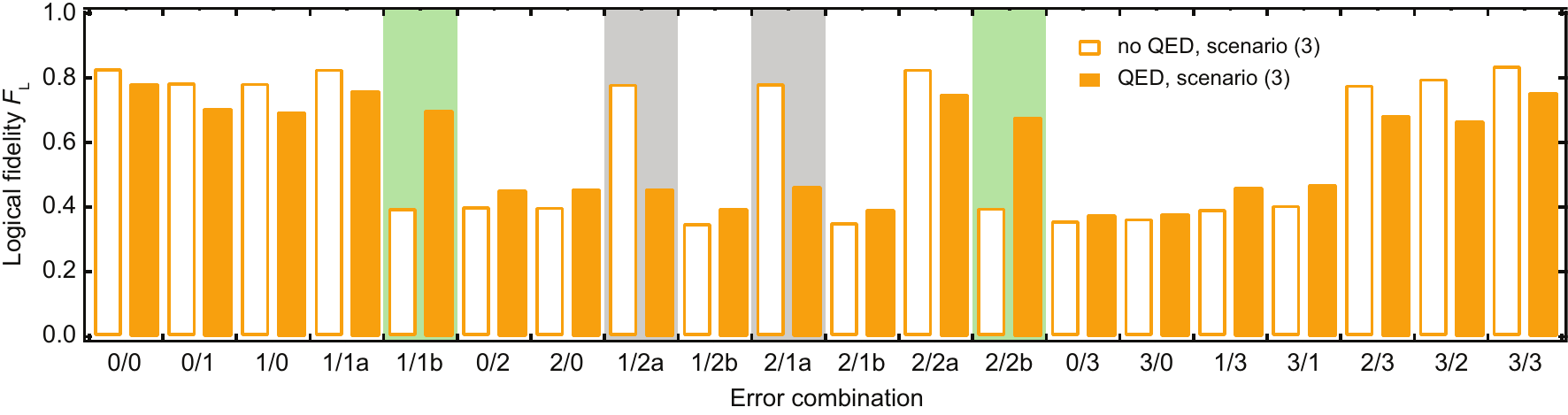}
\caption{\textbf{Comparison of logical fidelities $\FidL$ for all combinations of first- and second-round errors with and without  QED.} Same notation for error combinations as in Fig.~4d. Labels 1/1a and 2/2a indicate first- and second-round errors on the same qubits. Labels 1/2a and 2/1a indicate that one qubit undergoes errors in both rounds. 
	Green regions indicate the error combinations for which QED is expected to win over idling. Grey regions indicate the opposite. For all other combinations, QED and idling would ideally tie. }
\end{figure*}

\begin{figure*}[t]
\includegraphics[width=160mm]{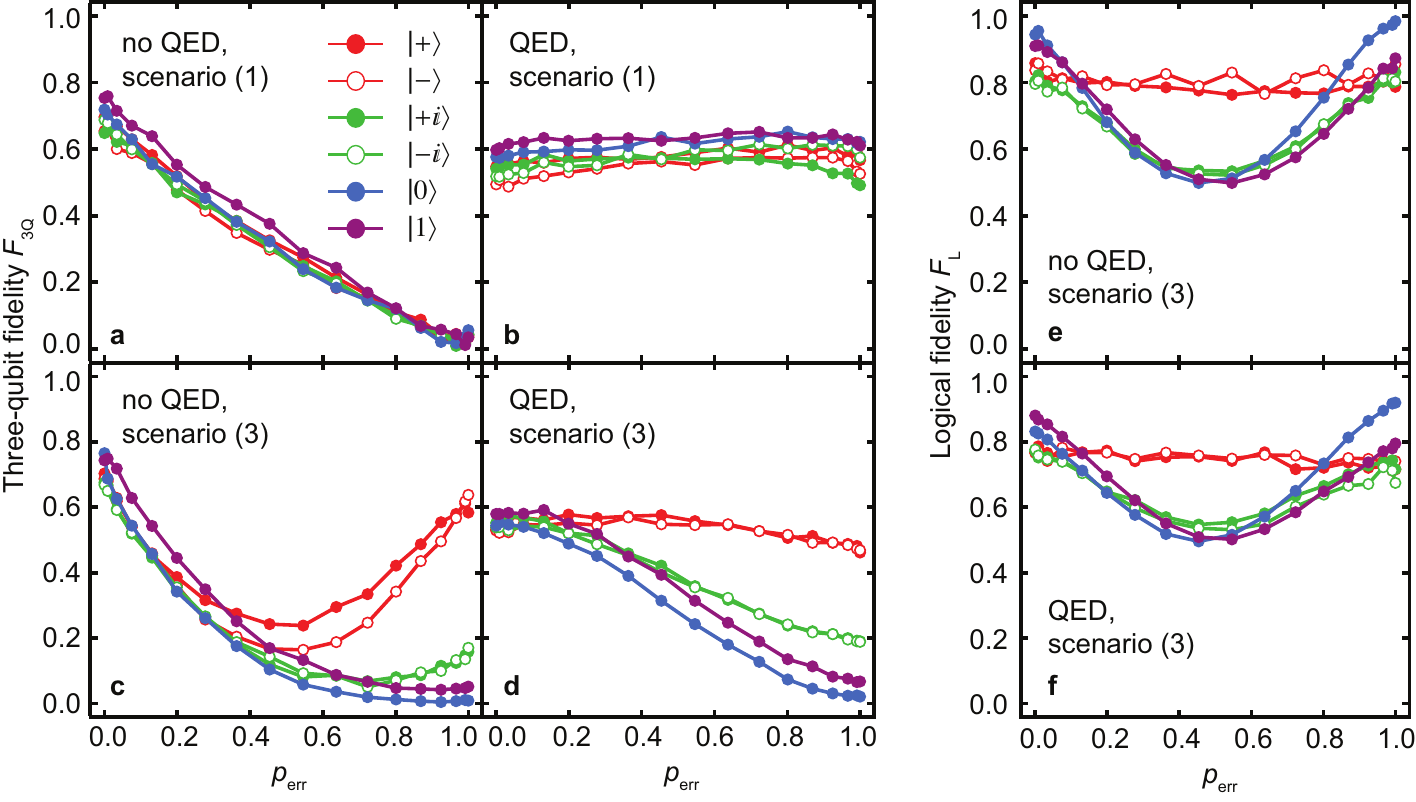}
\caption{\textbf{Three-qubit and logical state fidelities for the six cardinal input states of $\Dm$ under coherent bit-flip errors.}
	\textbf{a,b,} $\FidTQ$ for scenario (1) without and with QED, respectively. \textbf{c,d,} $\FidTQ$ for scenario (3) without and with QED, respectively. \textbf{e,f,} $\FidL$ for scenario (3) without and with QED, respectively.}
\end{figure*}

\begin{figure*} 
\includegraphics[width=2\columnwidth]{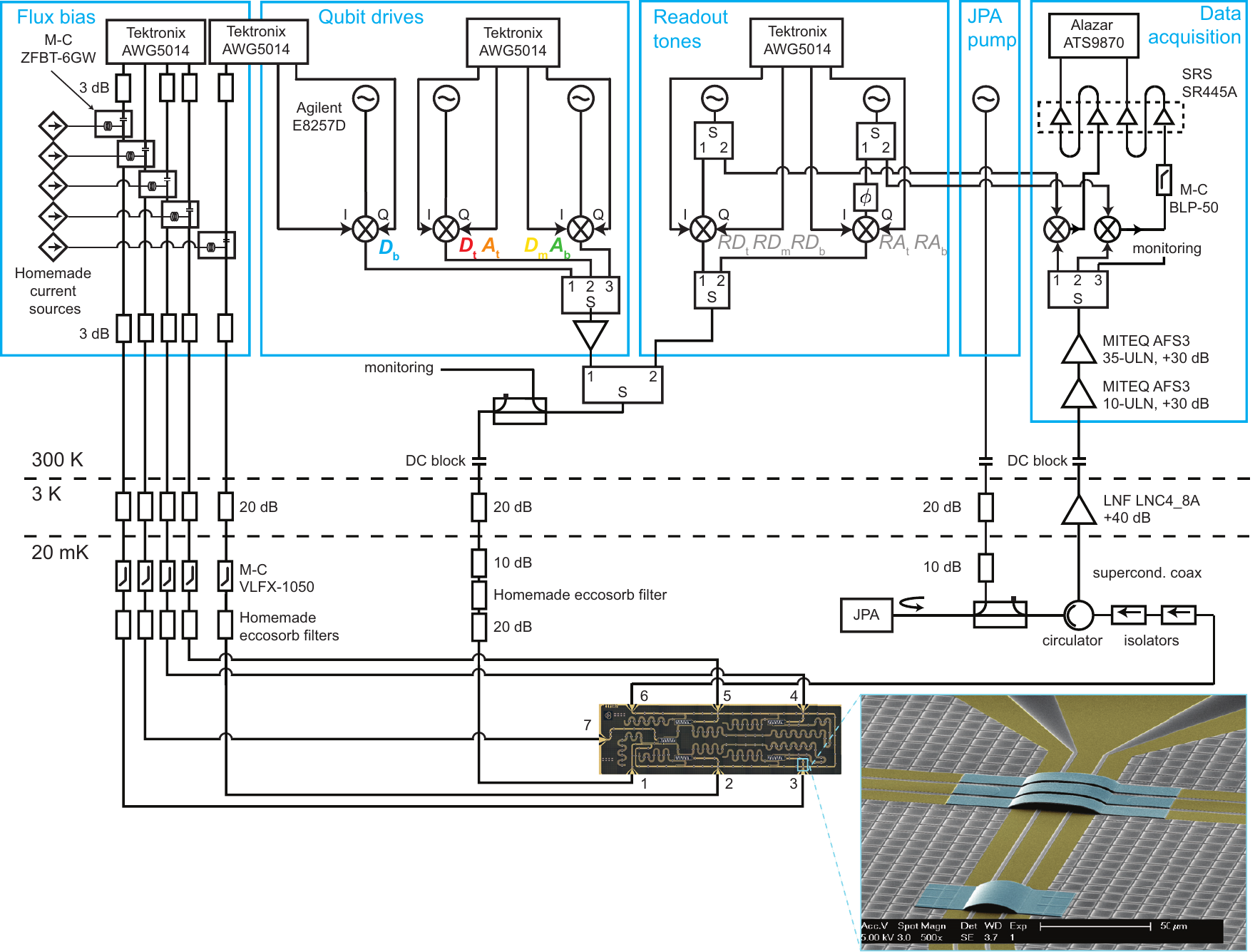}
\caption{\textbf{Experimental setup and device details.} Complete wiring of electronic components outside and inside the $\Hethree/\Hefour$ dilution refrigerator (Leiden Cryogenics CF-450). 
	Inset: False-color scanning electron micrograph showing processor details. Coplanar waveguide structures (resonators, feedline, and flux bias lines) are patterned on a NbTiN thin film (gold) on sapphire (gray). Al/Ti air bridges (blue) allow cross-overs between coplanar waveguide transmission lines, interconnections of ground planes, and suppression of slot-line mode propagation.}
\end{figure*}

\begin{figure*}
\includegraphics[width=\columnwidth]{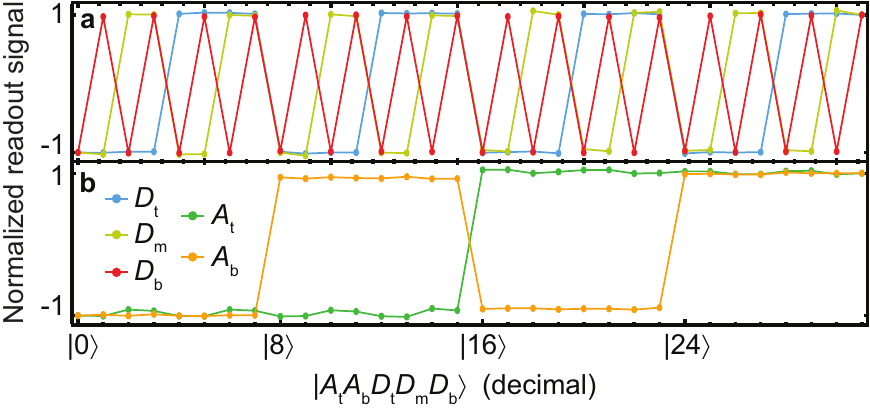}
\caption{\textbf{Low-crosstalk simultaneous qubit readouts.}
	Averaged and normalized readouts of the data (\textbf{a}) and ancilla (\textbf{b}) qubits immediately after preparing the five qubits in the 32 combinations of $\ket{0}$ and $\ket{1}$.}
\end{figure*}

\end{document}